\documentclass[aip, amsmath,amssymb,reprint]{revtex4-1}
\usepackage[english]{babel}
\usepackage{graphicx}
\usepackage{dcolumn}
\usepackage{bm}
\usepackage[colorlinks=true, allcolors=blue]{hyperref}
\usepackage{url}            
\usepackage{booktabs}       
\usepackage{amsfonts}       
\usepackage{nicefrac}       
\usepackage{microtype}      
\usepackage{lipsum}
\usepackage{fancyhdr}       
\usepackage[utf8]{inputenc}
\usepackage[T1]{fontenc}
\usepackage{mathptmx}
\usepackage{etoolbox}
\usepackage{svg}
\usepackage[normalem]{ulem}
\usepackage{dutchcal}
\usepackage{color}
\usepackage{xfrac}

\usepackage{xcolor} 

\makeatletter
\def\@email#1#2{%
 \endgroup
 \patchcmd{\titleblock@produce}
  {\frontmatter@RRAPformat}
  {\frontmatter@RRAPformat{\produce@RRAP{*#1\href{mailto:#2}{#2}}}\frontmatter@RRAPformat}
  {}{}
}%
\makeatother

\draft 

\begin{document}

\title{Multiscale wave-coupled neural networks for the inverse-design of broadband diffractive optics}



\author{Peyton Brown}
\affiliation{Department of Electrical \& Computer Engineering and Photonics Center, Boston University, 8 Saint Mary’s Street, Boston, Massachusetts 02215, USA}

\author{Riccardo Franchi}
\affiliation{Department of Electrical \& Computer Engineering and Photonics Center, Boston University, 8 Saint Mary’s Street, Boston, Massachusetts 02215, USA}

\author{Roberto Riganti}
\affiliation{Department of Physics, Boston University, 590 Commonwealth Avenue, Boston, Massachusetts 02215, USA}

\author{Jonathan Schuster}
\affiliation{Department of Electrical \& Computer Engineering and Photonics Center, Boston University, 8 Saint Mary’s Street, Boston, Massachusetts 02215, USA}
\affiliation{U.S Army Combat Capabilities Development Command (DEVCOM) Army Research Laboratory (ARL), 2800 Powder Mill Road, Adelphi, MD 20783-7291 USA}

\author{Luca Dal Negro$^*$}
\email{dalnegro@bu.edu}
\affiliation{Department of Electrical \& Computer Engineering and Photonics Center, Boston University, 8 Saint Mary’s Street, Boston, Massachusetts 02215, USA}
\affiliation{Department of Physics, Boston University, 590 Commonwealth Avenue, Boston, Massachusetts 02215, USA}
\affiliation{Division of Materials Science and Engineering, Boston University, 15 Saint Mary’s Street, Brookline, Massachusetts 02446,USA}

\date{July 13, 2026}

\begin{abstract}
In this work, we introduce an inverse-design model called multiscale wave-coupled neural networks (MSWNets) inspired by physics-informed neural networks (PINNs) that integrate the deep-learning capability of fully-connected neural networks with rigorous Rayleigh-Sommerfeld diffraction theory. MSWNets are trained using spatial input and output functions describing the height profiles of diffractive structures. Importantly, a MSWNet is composed of multiple sub-networks that provide learning across multiple frequency components of the optimal design solution. First, we validate and analyze the inverse design capabilities of the MSWNets' architecture through training on the complete complex field propagated from the analytical design of an axilens, demonstrating accurate reconstruction. Finally, we employ MSWNets, trained to maximize focusing efficiency, for the inverse design of novel diffractive devices. In particular, we design and display the performance of several achromatic lenses and doublets across the mid-wave infrared region ($3-5\rm{\;\mu} m$) with focal distance $f=500\rm{\;\mu m}$, demonstrating the potential of our inverse-design framework for the next-generation of broadband diffractive optical devices.
\end{abstract}


\maketitle 

\section{Introduction}
The increasing demand for compact, lightweight and easily integrable optical systems is driving a shift from traditional bulky refractive components to flat optics. In particular, diffractive optical elements (DOEs) and metasurfaces have received significant attention for the ability to engineer optical wavefronts over wavelength or subwavelength-scale thicknesses and facilitate imaging functionalities for miniaturized cameras and optical systems \cite{khonina_exploring_2024}. Within these broader categories, multilevel diffractive lenses (MDLs) and metalenses specifically represent the flat optics alternatives to conventional refractive lenses\cite{banerji_imaging_2019}. However, as required functionalities become increasingly sophisticated, traditional analytical methods have become impractical and inverse design algorithms have been demonstrated for the engineering of flat-optical devices with optimized performances. Specifically, inverse design algorithms have been utilized for the realization of ultracompact spectrometers \cite{zhu_design_2022}, lenses with an extended depth of focus (EDOF) \cite{banerji_extreme-depth--focus_2020, hayward_multilevel_2023, hayward_imaging_2024, bayati_inverse_2020, bayati_inverse_2022}, multi-focal optical devices \cite{chen_inverse_2022}, achromatic \cite{chung_high-na_2020, meem_inverse-designed_2020, meem_imaging_2021, jiang_multiwavelength_2023, xu_inverse_2025, maher_all_2023} and wide field-of-view (FOV)\cite{lin_inverse-designed_2023, hayward_multilevel_2023, arbabi_miniature_2016} lenses, and all-optical diffractive deep neural networks (D$^2$NNs) \cite{lin_all-optical_2018, luo_design_2019}.

Motivated by the recent advancements of physics-informed neural networks (PINNs) for the inverse design of optical materials~\cite{riganti_multiscale_2025}, this paper introduces multiscale wave-coupled neural networks (MSWNets), a hybrid inverse design framework that couples the deep-learning capabilities of neural networks with scalar diffraction theory for the inverse design of diffractive elements. To demonstrate the inverse-design capabilities, we first validate the framework by reconstructing the analytical profile of an axilens using a field-based optimization approach. Next, we leverage MSWNets to design broadband diffractive components using intensity-based training to maximize focusing efficiency. Specifically, we demonstrate this capability by first inverse-designing single-layer broadband achromatic flat lenses that maintain uniform focusing efficiency onto a single focal plane across the mid-wave infrared region ($3-5\rm{\;\mu} m$). Then, to overcome the performance limitations of the single-layer diffractive surface, we extend the framework to the inverse design of a dual-layer achromatic lens, referred to as a doublet, yielding superior optical performance. Finally, we assess the capabilities of these inverse-designed elements against standard benchmarks, including conventional Fresnel lenses and axilenses.
Ultimately, this work introduces a powerful neural network architecture capable of the simultaneous inverse design of multi-layer optical devices, enabling development of complex diffractive optical elements with desired performance.

\section{Multiscale wave-coupled neural networks}
\begin{figure*}
\includegraphics[width=0.95\textwidth]{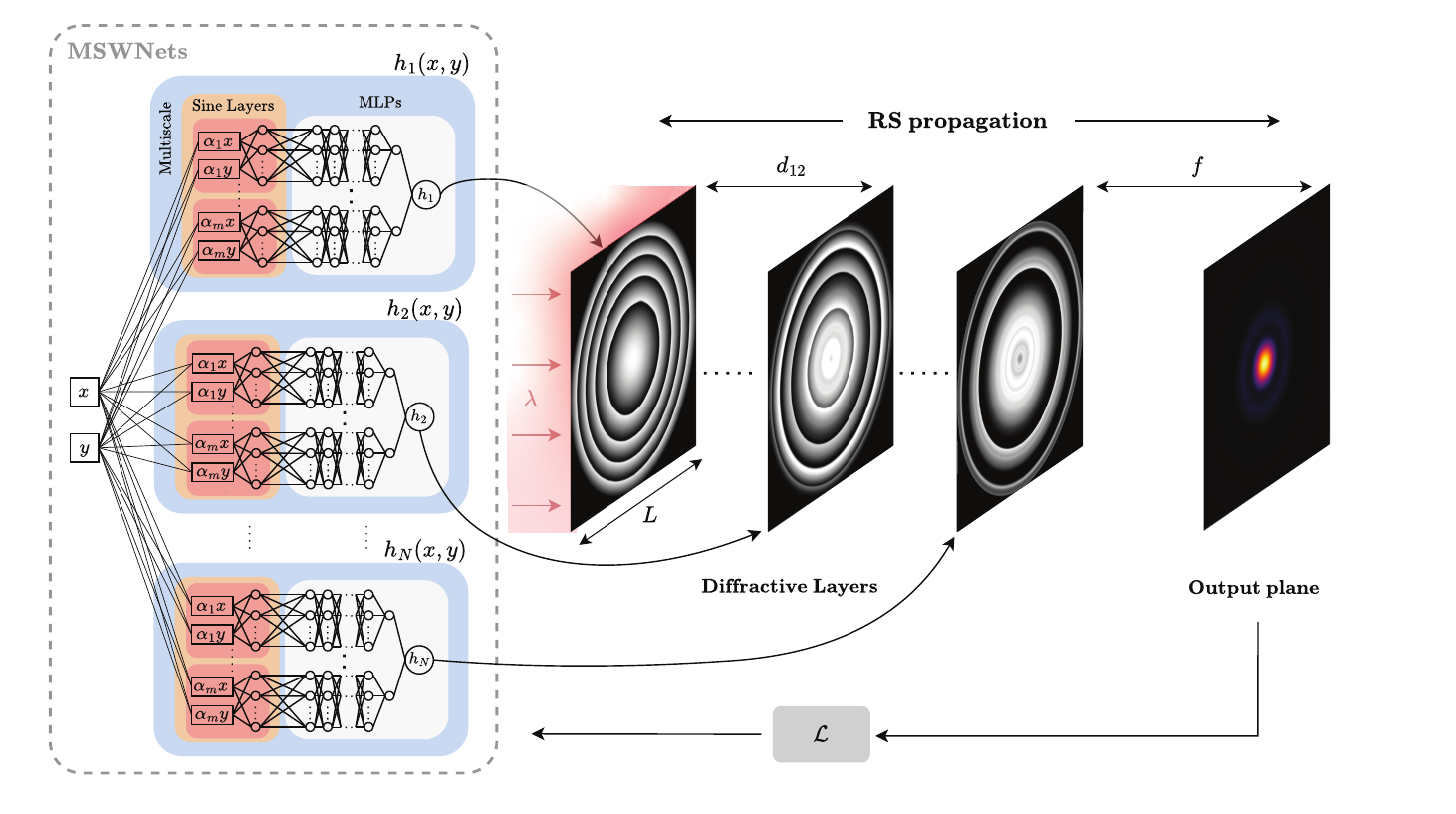}
\caption{Diagram of the multiscale wave-coupled neural
networks (MSWNet) architectures. The spatial design inputs are initially rescaled by $m$ multiscale sine features of scale $\alpha_{1,\dots,m}$, with the purpose of learning different components of the frequency domain of the device solution. The multiscale architectures then perform a forward pass on the rescaled inputs and linearly recombine the subnetwork outputs, applying a sigmoid function to bound the resulting diffractive layer $h_i$ within the $[0,h_{max}]$ range. Light incident on the first layer is propagated through the diffractive system using RS formulation and the network parameters are updated via gradient descent to optimize target optical functions.}
\label{fig:fig_1_wavenet}
\end{figure*}
An inverse-design problem seeks to recover the optimal topology or material distribution of a device $D$ within a defined computational domain $\Omega \subset \mathbb{R}^N$ that minimizes a specific objective function $\mathcal{L}$, subject to the governing physical constraints. In this work, we consider a general optical device $D$ composed of $N$ optical layers $h_i\in\mathbb{R}^2$, where $i\in\{1,N\}$. Each layer is a diffractive height profile of arbitrary dispersion with finite-size pixels at each position $(x_i,y_i)\in\Omega$. Consequently, each pixel's height determines the phase change of light going through it.

Mathematically, each layer can be represented by the function map $f_{i}(x,y)\rightarrow h_i$, reducing the inverse design problem to the functional optimization of $f_{i}(x,y)=h_{i,\Theta}(x,y)$ such that the general inverse design objective function is minimized:
\begin{equation}
\mathcal{L}\left[u(h_{1,\Theta}(x,y) \,, \dots,\,h_{N,\Theta}(x,y)), h_{1,\Theta}(x,y), \dots,h_{N,\Theta}(x,y)\right]
\label{eq:loss_general}
\end{equation}
Where $u$ is the function that propagates the field from the input plane to the output plane.
Here, $h_{i,\Theta}(x,y)$ are Multiscale Wave-Coupled Neural Networks shown schematically in Fig.~\ref{fig:fig_1_wavenet}. This diagram displays $N$ MSWNets instantiated from the desired number of diffractive layers of the optical device $D$ to be inversely designed. The architecture of each MSWNet is comprised of $m$ multiscale sub-networks that enable learning across the full frequency spectrum of the unknown optimal device. This multiscale feature is sometimes crucial for the network to capture the full frequency spectrum of the functional representation of the device, as shown by previous work~\cite{liu_multi-scale_2020, zhang_correction_2023, riganti_multiscale_2025}. Each sub-network is composed of an initial multiscale layer with the sine activation function and trainable weights and biases. The hidden layers have the sigmoid linear unit (SiLU, commonly known as Swish) activation function and the outputs are linearly combined to obtain the diffractive layer $h_{i,\Theta}(x,y)$.

To train the weights and biases $\Theta$ of the MSWNets, they are coupled to rigorous optical diffraction theory using the Rayleigh-Sommerfeld (RS) integral formulation. The field, $E(x',y';z)$, at a distance $z$, is obtained through the model\cite{goodman2017introduction}:
\begin{gather}
    \mathcal{h}(x,y;z,k) = \frac{z}{2\pi r^2}\left(\frac{1}{r}- jk \right)e^{ikr}, 
    \label{eq:impulse}
    \\
    E(x',y';z) = E(x,y;0) * \mathcal{h}(x,y;z,k),
    \label{eq:convolution}
\end{gather}
Here, $*$ denotes the two-dimensional spatial convolution, $k=2\pi n/\lambda$ is the wavenumber, $\lambda$ is the incident wavelength in vacuum, and $n$ is the index of the medium between the propagation planes. Additionally, $r=\sqrt{x^2 +y^2 + z^2}$, where $z$ is the distance between the planes. To accurately simulate light propagation between planes within a computational window of side $L$ and grid spacing $\delta x$, the impulse response is used when the propagation distance satisfies $z > L \sqrt{\left(2\delta x / \lambda \right) ^2 - 1}$. For all other distances, the transfer function (i.e., the Fourier transform of the impulse response) is used \cite{mehrabkhani_is_2017}.
Finally, an arbitrary optical objective is encoded into a loss function calculated at each forward simulation, which updates the weights of biases of the MSWNets via gradient descent using the Adam algorithm to optimize the height function $h(x,y)$ of the diffractive layers.
Then, to enforce fabrication constraints, we restricted the height profile from $0$ to $h_{max}$ using a sigmoid activation function at the network's final output.
Before each propagation step, the height values are converted into a phase profile at the incident wavelength under plane-wave illumination. The computational window is then zero-padded to improve the resolution and ensure the accuracy of diffraction calculations.
This method produces desired height profiles for each diffractive layer, allowing for multispectral optimization and the direct inverse-design of physical diffractive layers. This optimization process may produce rapidly-varying height profiles that greatly fluctuate pixel-to-pixel. Therefore, the number of trainable pixels is restricted to satisfy the Nyquist criteria ($f_s \geq 2f_{max}$) and pixels are added via linear interpolation to create the full mesh, ensuring proper sampling for propagation. This technique inherently produces smoother height profiles for fabrication feasibility and direct control of physical feature size. For objectives incorporating radially symmetric target intensities, the trained values are restricted to a vector, $r\in[0, L/2]$, and the resulting height vector is azimuthally revolved to generate the complete 2D height distribution, reducing computational demands. Additionally, to enhance numerical stability and accelerate training convergence, the radial coordinates are normalized to the $[0,1]$ interval prior to the forward pass through the neural network.

\section{Results and discussion}
\subsection{Axilens reconstruction based on complex fields}
First, we assess the inverse-design capabilities of our model through training a MSWNet on the field produced by an analytically-designed axilens to reconstruct its phase profile. An axilens is a diffractive optical element combining the point-focusing properties of a lens with the extended depth of focusing properties of an axicon, defined by the phase profile \cite{davidson_holographic_1991}
\begin{equation*}
    \phi_{axi}(r) = -\frac{2\pi}{\lambda_0} \left[ \sqrt{\left( f_0 + \frac{r \Delta f}{R} \right)^2 + r^2} - \left( f_0 + \frac{r \Delta f}{R} \right) \right]\left. \rule{0pt}{2.5em} \right|_{2\pi}
\end{equation*}
where $r=\sqrt{x^2+y^2}$, $R$ is the radius of the device aperture, $\lambda_0$ is the design wavelength, and $f_0$ and $\Delta f$ define the focal distance and depth. The phase can be converted to a thickness profile at a chosen design wavelength through $h(r) = \phi(r) \left( \frac{\lambda_0}{2\pi} \frac{1}{\Delta n} \right)$ where $\Delta n$ is the change in refractive index between the incident and DOE media. We design an axilens with the following parameters: $f_0=250 \rm{\;\mu m}$, $\Delta f = 100 \rm{\;\mu m}$, $R=102.4 \rm{\;\mu m}$, $n=3$, and $\lambda_0=3 \rm{\;\mu m}$. Using plane-wave excitation, light incident on the axilens is propagated using the RS formulation to obtain the field $E_{\rm{target}}$ at a distance $z$, which is incorporated into the loss function as:
\begin{equation}
    \mathcal{L} = \frac{1}{N} \left\| \tilde{E} - E_{target} \right\|_2^2
\end{equation}
where $\tilde{E}$ is the field propagated from the MSWNet height profile at each iteration, and $N$ is the number of samples in the simulation grid.
\begin{figure}
\includegraphics[width=\columnwidth]{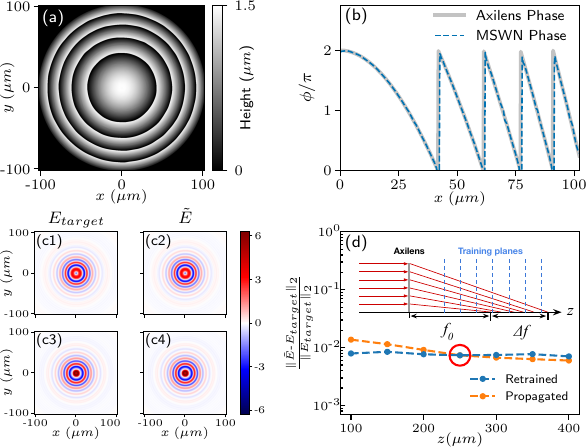}
\caption{(a) MSWNet-retrieved height profile training on the axilens field at $z=\rm{250 \ \mu m}$. (b) Line-cut of the retrieved phase (dashed blue) against the analytical axilens phase profile (solid gray). (c1) Real and (c3) imaginary components of the axilens training field $E_{target}$ and the (c2) real component and (c4) imaginary component of the fields propagated from the MSWN-retrieved phase $\tilde{E}$. (d) The relative $L_2$ error between $\tilde{E}$ and $E_{target}$ trained at each propagation distance (blue) and trained at one distance, $z=250 \; \rm{\mu m}$, and propagated to each plane (orange). The red circle indicates the training distance for (a)--(c).}
\label{fig:fig_2_axilens}
\end{figure}
To reconstruct the analytical phase profile, we train a single-scale MSWNet, comprising 4 hidden layers of 128 neurons each, and scale value of 102, using the target fields produced by the axilens.
Unlike the radially symmetric optimization previously described, the MSWNet is trained to retrieve the full grid $(x,y)$ of the diffractive layer with no interpolation for better network assessment. Fig.~\ref{fig:fig_2_axilens}(a) displays the height profile $h(x,y)$ obtained by MSWNet using a propagation distance of $z= \rm{250 \; \mu m}$.
A radial line-cut of the retrieved phase profile $\phi(x,0)$ (blue) is compared to the analytical axilens design (gray) in Fig.~\ref{fig:fig_2_axilens}(b), revealing excellent overlap. The field components at $z = 250\mathrm{\;\mu m}$ are shown in Fig.~\ref{fig:fig_2_axilens}(c). The left column displays the real (c1) and imaginary (c3) target fields produced by the axilens, while the right column presents the corresponding fields (c2, c4) produced from the network-retrieved profile, demonstrating excellent agreement.

To verify stability across different axial training planes, MSWNets were trained for fields produced by the axilens at distances $z=\rm{100 \; \mu m}$ to $z=\rm{400 \; \mu m}$ in steps of $50 \rm{\;\mu m}$. After each training, the accuracy of the MSWNet-designed DOE was quantified using the relative $L_2$ error between the training field and the retrieved field. Figure~\ref{fig:fig_2_axilens}(d) plots the relative $L_2$ error between the target and retrieved fields for a MSWNet retrained at each distance (blue), and the relative $L_2$ error between the target fields and a MSWNet trained at a single distance of $z=250 \rm{\;\mu m}$, and propagated to each plane (orange). These results show the stability of the model to accurately retrieve the axilens thickness distribution directly from field data. At each distance, the network was trained for 5000 epochs on a NVIDIA RTX Pro 6000 Blackwell GPU, with a typical training time of approximately 7 minutes. In the next sections, we extend the use of MSWNet for the optimization of intensity-based optical objectives that are not easily realized through traditional diffractive optics techniques.

\subsection{Achromatic lenses across the mid-infrared}
Having validated MSWNets using complex-field training for the reconstruction of an axilens, we now demonstrate the functional inverse-design capabilities of the network by optimizing diffractive profiles designed to focus light across well-separated wavelengths to a common focal plane. Specifically, we target the mid-wave infrared (MWIR) spectrum spanning $3-5 \rm{\;\mu m}$, a band which holds significance across diverse technologies, including thermal imaging, molecular fingerprinting, chemical spectroscopy, environmental monitoring, and astronomy \cite{tan_emerging_2018, argirusis_ir_2025, butt_mid-infrared_2026}. Conventional analytical designs, such as standard Fresnel lenses, suffer from severe chromatic aberrations that manifest as large axial focal shifts when the illumination wavelength deviates from its nominal value. A prominent framework for mitigating these aberrations is the harmonic diffractive lens \cite{sweeney_harmonic_1995}, which leverages an increased structural height to exploit multiple diffraction orders, enabling a common diffraction-limited focus at discrete, widely separated wavelengths. In these designs, the thickness is scaled by an integer multiple $m$ (the harmonic order) relative to a conventional modulo-$2\pi$ diffraction lens. Although a larger value of $m$ reduces the axial focal shift at intermediate wavelengths, the resulting structure becomes substantially thicker, exhibiting hybrid refractive and diffractive properties \cite{sweeney_harmonic_1995}. While Engelberg and Levy provide a thorough study on the performance and limitations of such harmonic configurations \cite{engelberg_achromatic_2021}, a detailed analysis of these architectures is left outside the scope of this paper to maintain our focus on ultra-thin lenses operating strictly in the diffractive regime.

Achieving chromatic correction while maintaining an ultra-thin profile remains a significant challenge that traditional analytical configurations cannot easily resolve. To address this, we leverage MSWNet optimization to inverse-design several diffractive flat lenses and doublets. We evaluate designs with maximum phase shifts, $\Delta \phi_{max}$, of $2\pi$, $4\pi$, and $6\pi$, to systematically showcase the tradeoff between structural height and broadband achromatic performance. Notably, our optimization targets a uniform response across the entire $3-5 \rm{\;\mu m}$ spectrum to ensure consistent focusing performance across all operating wavelengths.

\subsubsection{Intensity-based training scheme for achromatic performance}
Because standard imaging applications rely solely on the intensity of the focused light, the phase at the focal plane can be neglected, shifting our training to an entirely intensity-based optimization scheme. Here, we define the network efficiencies $\tilde{\eta}_i$ corresponding to each target wavelength $\lambda_i$ focused to the designed focal plane $z=f$ as
\begin{equation}
    \tilde{\eta}_i = \frac{\int_0^{2\pi} \int_0^{R_\eta}I'(\lambda_i, z=f, \rho', \theta')dr'd\theta'}{\int \int I(\lambda_i, z=0, \rho,\theta)dS}
\end{equation}
where $I'$ denotes the intensity distribution on the focal plane, $I$ denotes the intensity distribution on the input plane, and $S$ denotes the input aperture. The symbols $(\rho,\theta)$ and $(\rho',\theta')$ denote the polar coordinates on the input and output plane, respectively. The integration radius $R_\eta$ represents the desired radius of the focused intensity. To improve the intensity profile at the focal plane during network training, we set $R_\eta$ equal to the diffraction-limited full-width at half maximum ($\mathrm{FWHM} = 0.514 \lambda / \mathrm{NA}$) for each wavelength within the target spectral band, rather than the conventional $1.5\times\mathrm{FWHM}$ typically used to calculate focusing efficiency \cite{meem_full-color_2018, meem_broadband_2019, balli_hybrid_2020, bayati_inverse_2022, bayati_inverse_2020,xiong_polarization-independent_2022}. Here, NA refers to the numerical aperture of the lens given by $\mathrm{NA} = n\sin(\tan^{-1}(L / 2f))$ where $n$ here is the refractive index of the final propagation media to the focal plane at distance $f$ and $L$ is the diameter of the entrance pupil. In this optimization process, conventional mean-squared-error (MSE) loss functions may cause the optimization to favor wavelengths that are easier to focus. To obtain a flat, broadband response across the target spectral band, we use a composite loss function incorporating a Log-Sum-Exp (LSE) term and a variance penalty to specifically target the wavelengths with the lowest efficiency \cite{maher_all_2023, maher_highly_2025}:
\begin{equation}
    \mathcal{L} = \frac{1}{\beta}\log\left(\sum_{i=1}^N\exp\left( \beta(1-\tilde{\eta}_i) \right) \right) +w_v\sigma_\eta^2
\end{equation}
The first term serves as a smooth approximation of the maximum loss, where the parameter $\beta$ controls the selectivity of the targeted correction.
For very large values of $\beta$, the network optimizes only the single worst-performing wavelength at any given step, which often leads to unstable and discontinuous training as the targeted wavelength rapidly shifts. Conversely, small values of $\beta$ dilute the focus across all wavelengths simultaneously. Therefore, we operate in an intermediate $\beta$ regime: this compels the network to continuously focus on a subset of wavelengths exhibiting the lowest efficiencies, preventing erratic focal shifts and ensuring a smoother, more stable optimization process.
Furthermore, the additional variance penalty (scaled by weight $w_v$) further enforces a uniform response across the entire target spectrum. Because we observed that a strictly dominant variance penalty suppresses the average efficiency, $w_v$ was empirically tuned to establish an optimal compromise between overall efficiency and flat multispectral performance.

\begin{figure}[t]
\includegraphics[width=\columnwidth]{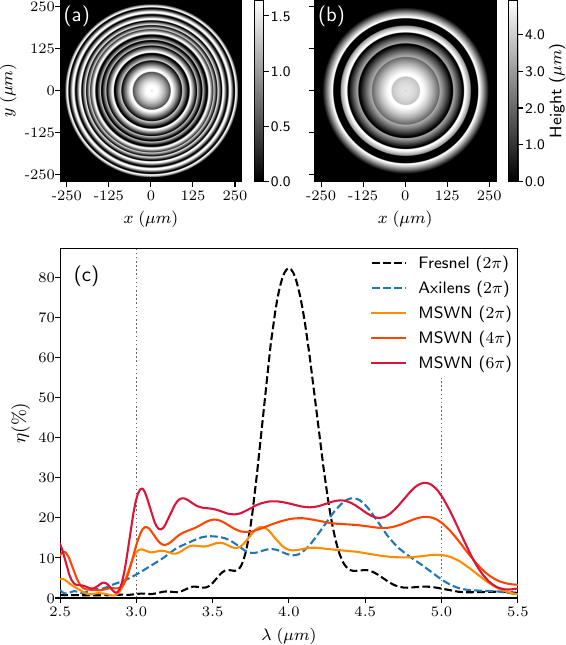}
\caption{MSWN-retrieved height profiles of the single-layer achromatic lenses utilizing (a) a maximum height of 1.64 $\rm{\mu m}$ corresponding to a maximum phase shift of $2\pi$ at $\lambda=4 \rm{\;\mu m}$ and (b) a maximum height of 4.93 $\rm{\mu m}$ corresponding to a maximum phase shift of $6\pi$ at $\lambda=4 \rm{\;\mu m}$. (c) Simulated focusing efficiencies $\eta$ as a function of wavelength for the MSWN-retrieved lenses using max heights determined by a $2\pi$, $4\pi$, and $6\pi$ phase shift at $\lambda=4 \rm{\;\mu m}$ (yellow, orange, and red respectively), axilens (dashed blue), and Fresnel lens (dashed black). The target spectral range ($3-5\rm{\; \mu m}$) is indicated by the vertical gray lines.}
\label{fig:achromatic_1layer}
\end{figure}

\subsubsection{Single-layer achromatic lenses}
Implementing this training scheme, we first employ MSWNets for the optimization and design of single diffractive structures with increasing maximum heights of 1.64, 3.29, and 4.93 $\rm{\mu m}$, corresponding to maximum phase shifts $\Delta \phi_{max}$ of $2\pi$, $4\pi$, $6\pi$ at the central wavelength $\lambda_{c}=4 \rm{\;\mu m}$. We assume the dispersion of silicon, which has a refractive index of $n=3.45$ to $3.42$ across the spectral range, and is compatible with standard complementary metal-oxide-semiconductor (CMOS) fabrication processes. Next we initialize diffractive layers with side length $L=512 \rm{\;\mu m}$ and focal distance $f=500\rm{\;\mu m}$ to design achromatic lenses with $\mathrm{NA}=0.45$ suitable for compact imaging applications, using a pixel size of $\delta x =1\rm{\;\mu m}$. MSWNets were trained on 21 discrete wavelengths from $3-5 \rm{\;\mu m}$ over 2000 epochs on a NVIDIA RTX 6000 Pro, taking around 6 minutes each. More information on the network parameters can be found in Sec. 1 of the supplementary material. The inverse-designed lenses using $h_{max}$ ($\Delta \phi_{max}$) of 1.64 ($2\pi$) and 4.93 $\rm{\mu m}$ ($6\pi$) are displayed in Fig. \ref{fig:achromatic_1layer}(a) and (b) respectively. As expected, the height profile corresponding to the $6\pi$ design is smoother than its $2\pi$ counterpart, featuring significantly fewer abrupt phase wraps from $h_{max}$ to $0$.

As introduced previously, an axilens is characterized by an extended focal depth which induces a spatial overlap of the focused intensities across a wide spectral band, inherently establishing an achromatic focal plane \cite{chen_phase-modulated_2020, britton_phase-modulated_2020, chen_design_2020}. As such, the axilens serves as a natural baseline for evaluating achromatic performance. For a comparison, we design an axilens using the parameters $f_0=300 \rm{\;\mu m}$, $\Delta f=200\rm{\;\mu m}$ and $\lambda=4\rm{\;\mu m}$, chosen for its best achromatic performance across the target band. We also design a Fresnel lens with $f=500\rm{\;\mu m}$ at the central wavelength $\lambda=4\rm{\;\mu m}$. We define the simulated focusing efficiency $\eta$ as the ratio of intensity within a radius of $1.5\times$ the diffraction-limited FWHM to the intensity incident on the (first) aperture. In this approach, the diffraction limited spot size is determined for each wavelength, resulting in an increasing integration radius for the efficiency calculation with increasing wavelength.

Figure \ref{fig:achromatic_1layer}(c) illustrates the simulated focusing efficiency of all the designs as a function of wavelength.
Characteristically, the Fresnel lens (black dashed line) exhibits a sharp efficiency peak at the design wavelength of $\lambda=4 \rm{\;\mu m}$, achieving the highest overall efficiency of $\approx82\%$ but a restricted bandwidth of about $0.65\rm{\;\mu m}$ (defined for $\eta>10\%$). Conversely, while the axilens provides an extended bandwidth, it suffers from a significantly reduced overall efficiency and a non uniform efficiency curve.
Our inverse-designed lenses address this fluctuating profile by maintaining a flat performance across the entire target spectrum, which yields slightly higher efficiencies near the band edges at $\lambda=3$ and $5\mathrm{\;\mu m}$.
Importantly, Fig. \ref{fig:achromatic_1layer}(c) highlights the effect of maximum height on diffractive designs, showing an increase in efficiency across the target band as we increase the allowed height of the lens.

\begin{figure}[t]
\includegraphics[width=\columnwidth]{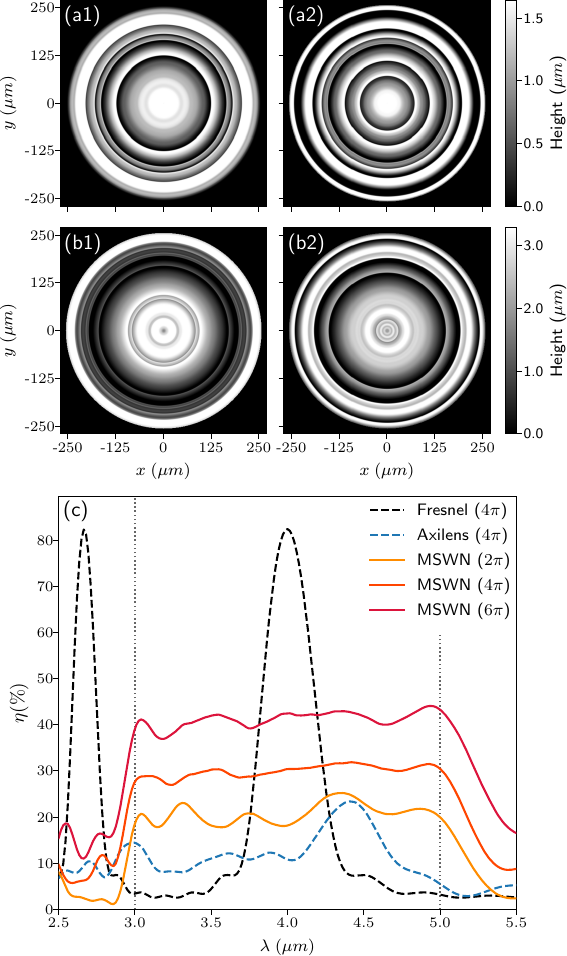}
\caption{MSWN-retrieved height profiles of the achromatic doublets utilizing (a1, a2) a maximum height of 1.64 $\rm{\mu m}$ corresponding to a maximum phase shift of $2\pi$ at $\lambda=4 \rm{\;\mu m}$ and (b1, b2) a maximum height of 3.29 $\rm{\mu m}$ corresponding to a maximum phase shift of $4\pi$ at $\lambda=4 \rm{\;\mu m}$. (c) Simulated focusing efficiencies $\eta$ as a function of wavelength of the MSWN-retrieved doublets using max heights determined by a $2\pi$, $4\pi$, and $6\pi$ phase shift at $\lambda=4 \rm{\;\mu m}$ (yellow, orange, and red respectively), axilens (dashed blue), and Fresnel lens (dashed black) designed using a maximum height corresponding to a $4\pi$ phase shift. A secondary peak of the Fresnel lens is seen around $\lambda=2.6\rm{\;\mu m}$ due to the third diffracted order of the Fresnel lens designed at $\lambda=2\times4 \rm{\;\mu m}$. The target spectral range ($3-5\rm{\; \mu m}$) is indicated by the vertical gray lines.}
\label{fig:achromatic_2layer}
\end{figure}

\subsubsection{Dual-layer achromatic doublets}
To further improve the achromatic performance, we introduce a second diffractive layer with similar maximum height constraints, utilizing doublet technology \cite{wirth-singh_wide_2025,lin_inverse-designed_2023, groever_meta-lens_2017, arbabi_miniature_2016}. For these designs, we select two diffractive layers with side lengths $L=512 \rm{\;\mu m}$ and inter-layer separation $d=500\rm{\;\mu m}$ representing the thickness of a standard silicon substrate, and a focal distance of $f=500\rm{\;\mu m}$ corresponding to the same numerical aperture ($\mathrm{NA}=0.45$) as the single-layer lenses. We choose the same maximum height constraints as the previous designs, and study the achromatic performance in relation to a Fresnel lens and axilens at increased maximum heights corresponding to $\Delta \phi_{max}=4\pi$. MSWNets were configured with two diffractive layers and trained on 21 discrete wavelengths between $3-5 \rm{\; \mu m}$ over 2000 epochs, taking around 11 minutes each.
Additional network parameters are reported in Sec. 1 of the supplementary material.
Figure \ref{fig:achromatic_2layer}(a1, a2) and \ref{fig:achromatic_2layer}(b1, b2) display the inverse-designed doublets for heights corresponding to a $2\pi$ and $4\pi$ phase shift, respectively.
Fig. \ref{fig:achromatic_2layer}(c) plots the simulated focusing efficiencies of the doublet designs, displaying a drastic improvement from the optimized single-layer designs at similar maximum heights. The inverse-designed doublet utilizing a maximum phase shift of $2\pi$ achieves an average focusing efficiency of $20.8\%$, while the single-layer counterpart reaches only $12.3\%$. The same doublet also displays improved performance over the axilens designed at an increased maximum height ($\phi_{\rm{max}}=4\pi$). Table \ref{tab:effs} lists the minimum, maximum, and average focusing efficiencies of each design using $\phi_{\rm{max}}=2\pi$, $4\pi$, and $6\pi$. While not displayed here, the efficiency curve of the Fresnel and axilens designed for $\phi_{\rm{max}}=6\pi$ is shown in Sec. 2 of the supplementary material.

While scaling the height of the designs improves achromatic performance, it requires balancing fabrication feasibility and the validity of scalar diffraction theory. Deeper structural features produce challenging high-aspect-ratio features and can induce resonance or shadowing effects that compromise the thin-element approximation (TEA), breaking the validity of scalar diffraction. To avoid this problem, local zone widths must equal or exceed the incident free-space wavelength \cite{mellin_limits_2001}. While all of our doublet designs comply with this requirement, maintaining a minimum zone width of 16 $\rm{\;\mu m}$, we choose to focus our primary analysis on the inverse-designed $4\pi$ doublet, which operates well within both scalar theory and current manufacturing limits. The $6\pi$ doublet with improved performance as featured in Fig. \ref{fig:achromatic_2layer}(c), is analyzed in Sec. 2 of the supplementary material, including its height profiles, focusing efficiencies, and intensity point spread functions (PSFs). A more detailed analysis of the designed structures regarding the validity of scalar diffraction theory can be found in Sec. 3 in the supplementary material.

\begin{table}
\caption{\label{tab:effs}
        Simulated focusing efficiencies in the $3-5\mathrm{\;\mu m}$ band.
        }
\begin{ruledtabular}
\begin{tabular}{cccc}
Design & $\rm{min(\eta)}$ (\%) & $\rm{max(\eta)}$ (\%) & $\overline{\eta}$ (\%)
\vspace{0.3cm} \\
Fresnel & &  & \\
\hline
$2\pi$ & 0.9 & 82.4 & 18.5\\
$4\pi$ & 2.5 & 82.4 & 19.1\\
$6\pi$ & 5.3 & 82.4 & 25.0\\

Axilens & &  & \\
\hline
$2\pi$ & 4.9 & 24.9 & 13.8\\
$4\pi$ & 5.8 & 23.4 & 12.6\\
$6\pi$ & 7.6 & 20.8 & 12.7\\

MSWN singlets & &  & \\
\hline
$2\pi$ & 10.1 & 17.7 & 12.3\\
$4\pi$ & 13.3 & 20.3 & 18.0\\
$6\pi$ & 17.1 & 28.8 & 23.2\\

MSWN doublets & & &\\
\hline
$2\pi$ & 17.8 & 25.2 & 20.8\\
$4\pi$ & 27.0 & 31.9 & 30.0\\
$6\pi$ & 36.9 & 44.0 & 41.2\\
\end{tabular}
\end{ruledtabular}
\end{table}

\begin{figure}[hbtp!]
\includegraphics[width=0.8\linewidth]{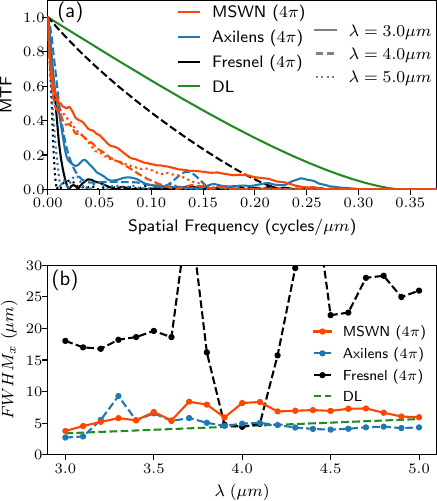}
\caption{Performance metrics of lens designs. (a) Modulation-transfer function (MTF) of the inverse-designed lens (orange) compared with the axilens (blue), Fresnel lens (black) and diffraction-limited (green) with wavelength indicated by the solid, dashed, and dotted lines. (b) Simulated full-width at half-maximum (FWHM) values for each design as a function of wavelength, compared against the theoretical diffraction limit.}
\label{fig:achromatic_2layer_MTFs_FWHMS}
\end{figure}

\begin{figure*}[hbtp!]
\includegraphics{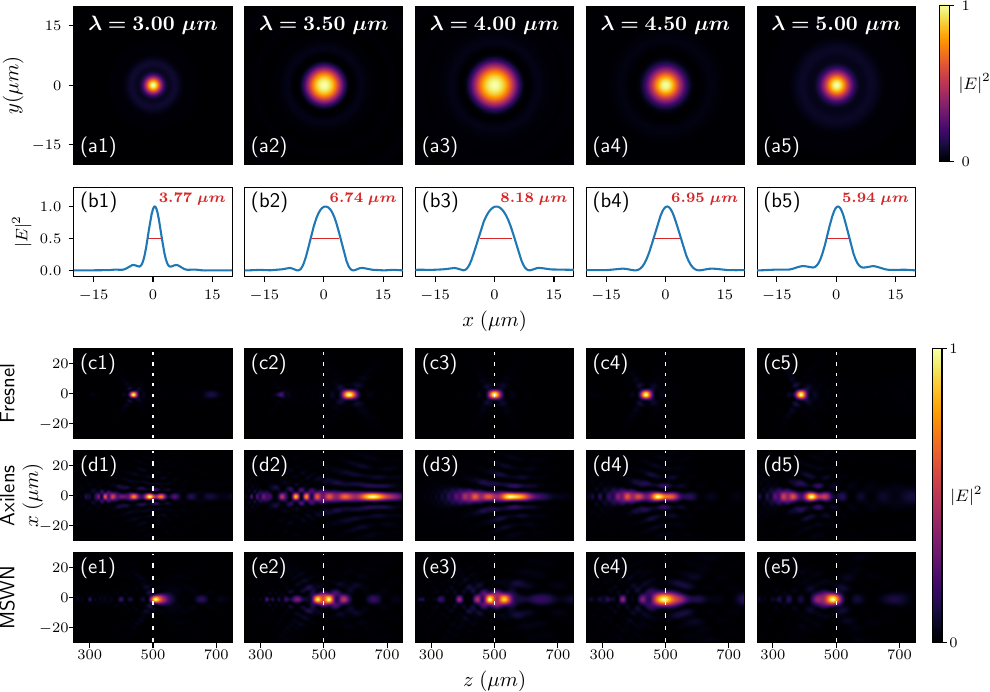}
\caption{Wavelength-dependent focal characterization of the inverse-designed achromatic lens with a maximum height corresponding to a $4\pi$ phase-shift at $\lambda=4\rm{\;\mu m}$. (a1-a5) Normalized 2D transverse intensity profiles $|E|^2$ of the inverse-designed achromatic doublet at the designed focal plane ($f=500 \rm{\;\mu m}$) across the $3-5 \rm{\;\mu m}$ wavelength spectrum. (b1-b5) Corresponding 1D horizontal line-cuts ($y=0$). Full-width at half-maximums (FWHMs) fit are indicated by the red bounds with values shown in the top right of each panel. (c1-c5) to (e1-e5) Cross-sectional longitudinal intensity profiles along the propagation direction for: (c1-c5) a conventional Fresnel lens, (d1-d5) an axilens, and (e1-e5) the MSWN inverse-designed doublet. The vertical dotted white lines indicate the target focal length ($f=500 \rm{\;\mu m}$). Intensity values for all 2D and longitudinal plots are independently normalized to their respective global maxima. The columns display the results for different wavelengths $\lambda \in \{3.0, 3.5, 4.0, 4.5, 5.0\}\mathrm{\;\mu m}$.}
\label{fig:achromatic_PSFs}
\end{figure*}

Figure \ref{fig:achromatic_2layer_MTFs_FWHMS}(a) displays the modulation transfer function (MTF) for each design at $\lambda \in \{3.0, 4.0, 5.0\}\mathrm{\;\mu m}$, quantifying the system's ability to image spatial frequencies.
The simulated MTF was calculated by taking the Fourier transform of the focal plane intensity PSF and normalizing it to its zero-frequency value.
At its design wavelength ($\lambda=4.0\mathrm{\;\mu m}$), the Fresnel lens exhibits the broadest spatial frequency response, but its performance degrades rapidly at off-target wavelengths. While the axilens preserves a moderate frequency transfer across the spectrum, the inverse-designed MSWN doublet consistently outperforms it, maintaining superior resolution across all three evaluated wavelengths. The diffraction-limited (DL) MTF at $\lambda=3.0\mathrm{\;\mu m}$ (solid green line) is included as a theoretical reference.
The FWHMs, obtained by fitting the focal intensity to an Airy pattern, are presented in Fig.~\ref{fig:achromatic_2layer_MTFs_FWHMS}(b).
While the FWHM of the Fresnel lens diverges rapidly when moving away from its $\lambda=4.0\mathrm{\;\mu m}$ design wavelength, both the inverse-designed doublet and the axilens exhibit highly stable spot sizes that remain near the theoretical diffraction limit across the spectrum.
Notably, the fitted axilens values (blue) fall artificially below the diffraction limit (green); this occurs because the numerical fit primarily captures the narrow central peak, neglecting the significant energy distributed across the surrounding side-lobes. Further details regarding the fitting procedure can be found in Sec. 4 of the supplementary material.

\begin{figure}[t]
\includegraphics[width=0.9\linewidth]{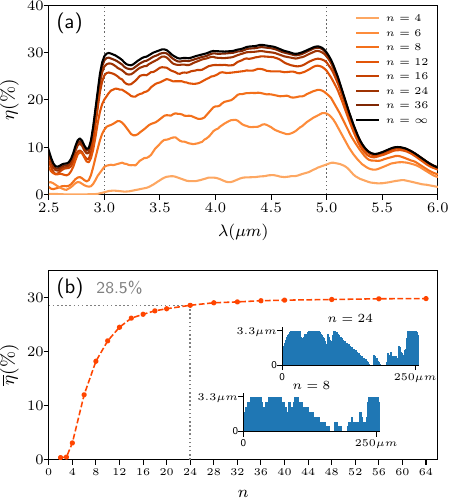}
\caption{(a) Simulated focusing efficiency of the inverse-designed achromatic doublet as a function of wavelength at various numbers of height discretization levels. (b) Average efficiency across the target spectral range ($3-5\rm{\;\mu m}$) as a function of discretized levels. The dotted gray line highlights the average focusing efficiency of $\overline{\eta}=28.5$ at only 24 height levels. Insets show the height distribution of the discretized height profiles at 8 and 24 levels for reference.}
\label{fig:achromatic_discretized}
\end{figure}

To further analyze the optical performance of the dual-layer inverse-designed doublet, Fig.~\ref{fig:achromatic_PSFs} compares its transverse and longitudinal intensity distributions across $\lambda \in \{3.0, 3.5, 4.0, 4.5, 5.0\}\mathrm{\;\mu m}$ against those of the conventional Fresnel lens and axilens.
Specifically, the transverse intensity profiles of the MSWN doublet at the designed focal plane ($f=500\mathrm{\;\mu m}$) are depicted in Fig.~\ref{fig:achromatic_PSFs}(a1)-(a5), with the corresponding 1D line-cuts displayed in Fig.~\ref{fig:achromatic_PSFs}(b1)-(b5).
The extracted FWHMs are indicated by the red horizontal lines, with the explicit values provided in the top right of each panel.
Furthermore, Fig.~\ref{fig:achromatic_PSFs}(c1)-(c5), (d1)-(d5), and (e1)-(e5) display the longitudinal intensity profiles of the focused intensity for the Fresnel lens, the axilens, and the inverse-designed doublet, respectively, where the white vertical lines denote the designed focal length.
Evidently, the Fresnel lens suffers from significant chromatic focal shift at the off-design wavelengths.
While the axilens leverages its extended depth of focus to maintain some achromaticity across the spectrum, it is ultimately outperformed by the inverse-designed doublet, which successfully maintains a high focused intensity at the target focal plane across the entire target wavelength band.

Traditionally, multilevel diffractive structures were fabricated using a hard-mask and $N$ number of exposure steps to create $2^N$ steps approximating a grayscale design. However, the past several years has seen major advancements in maskless photolithography and etching processes \cite{kim_optimization_2022, bouchouri_optimization_2024, lin_inverse-designed_2023, yoshie_bias-supply_2023, li_recent_2024}, eliminating the need for multiple photolithography steps and boosting large-scale manufacturing with a single grayscale exposure. In one implementation, Meem et al. inverse-design, fabricate and experimentally demonstrate an MDL designed using 100 height levels \cite{meem_imaging_2021}. The design was fabricated in photoresist atop glass and silicon substrates and utilized a maximum height of $10 \rm{\;\mu m}$, which the authors note introduces a phase shift much larger than $2\pi$ for many of the target wavelengths.
Metalenses, on the other hand, require subwavelength features increasing aspect ratios that require more complicated fabrication processes \cite{banerji_imaging_2019}, reducing throughput potential.
Nonetheless, we study the effects of post-training height discretization on the achromatic performance, as seen in Fig. \ref{fig:achromatic_discretized}(a), where we plot the simulated focusing efficiency of the inverse-designed achromatic doublet as a function of wavelength at various levels of discretization $n$.
Here, the efficiencies approach the black efficiency curve, representing the simulated efficiency of the grayscale ($n=\infty$) structure. Figure~\ref{fig:achromatic_discretized}(b) shows the average simulated focusing efficiency, $\overline{\eta}$, across the target spectrum as a function of the number of discretized height levels, $n$.
As the number of levels increases, the average efficiency grows rapidly at first, ultimately saturating and plateauing at $36$ levels and beyond. Notably, with only 24 discretized height levels, the achromatic doublet achieves an average focusing efficiency of $\approx 29\%$, as indicated by the dotted gray lines in Fig.~\ref{fig:achromatic_discretized}.

\section{Conclusions}
In conclusion, we have introduced MSWNet, an inverse design method inspired by PINNs, which incorporates a fully-connected deep neural network for the optimization and inverse-design of multiple diffractive layers given a desired optical task encoded through a loss function.
We detail our network and optimization framework, which integrates rigorous Rayleigh-Sommerfeld scalar diffraction theory for the inverse design of various diffractive optical elements.
The capabilities of the network were first validated through complex-field training targeting the reconstruction of an analytically designed axilens. Next, we demonstrated the practical utility of the model by designing and analyzing several broadband achromatic flat lenses and doublets operating across the $3-5 \rm{\;\mu m}$ mid-wave infrared spectrum with a focal distance of $f=500\rm{\;\mu m}$ and diameter $512\rm{\;\mu m}$ ($\mathrm{NA}=0.45$).
For imaging applications, we utilized an intensity-based optimization scheme to maximize the focusing efficiency across the target bandwidth.
By evaluating designs with maximum phase depths of $\Delta \phi_{max}=2\pi$, $4\pi$, and $6\pi$, we characterized the trade-off between height profiles and multispectral focusing performance.
We further analyze the doublet using $\Delta \phi_{max}=4\pi$, showcasing its performance metrics and PSFs in comparison to traditional diffractive devices such as the axilens and standard Fresnel lens.
The inverse-designed doublet achieves improved transverse and longitudinal intensity profiles, delivering an average efficiency of $30\%$ across the $3-5\mathrm{\;\mu m}$ wavelength range.
Ultimately, this work establishes MSWNets as a powerful tool for the optimization and inverse design of high-performance multi-layer diffractive optics for next-generation multifunctional and ultracompact optical systems.

\section*{Supplementary material}
See the supplementary material for additional information.

\begin{acknowledgments}
Distribution Statement A: Approved for public release: distribution is unlimited. Research was sponsored by the Army Research Laboratory and was accomplished under Cooperative Agreement Number W911NF-25-2-0135. The views and conclusions contained in this document are those of the authors and should not be interpreted as representing the official policies, either expressed or implied, of the Army Research Laboratory or the U.S. Government. The U.S. Government is authorized to reproduce and distribute reprints for Government purposes notwithstanding any copyright notation herein.
The authors are pleased to acknowledge that the computational work reported in this paper was performed on the Shared Computing Cluster, which is administered by Boston University’s Research Computing Services.

\end{acknowledgments}

\section*{Author declarations}
\subsection*{Conflict of Interest}
The authors have no conflicts to disclose.

\section*{Data Availability}
The data that support the findings of this study are available from the corresponding author upon reasonable request.

\bibliography{references}



\clearpage

\setcounter{section}{0}  
\setcounter{figure}{0}   
\setcounter{table}{0}    
\setcounter{equation}{0} 

\renewcommand{\thefigure}{S\arabic{figure}}     
\renewcommand{\thetable}{S\arabic{table}}       

\renewcommand{\theequation}{S\arabic{equation}} 
\renewcommand{\thesection}{S\arabic{section}}   









\onecolumngrid 
\vspace*{1cm}
\begin{center}
    \textbf{\Large Supplementary Material: Multiscale wave-coupled neural networks for the inverse-design of broadband diffractive optics}\\[0.6cm]
    
    Peyton Brown$^{1}$, Riccardo Franchi$^{1}$, Roberto Riganti$^{2}$, Jonathan Schuster$^{1, 3}$, and Luca Dal Negro$^{1, 2, 4, *}$\\[0.4cm]
    
    {\small \itshape
    $^1$Department of Electrical \& Computer Engineering and Photonics Center, Boston University, 8 Saint Mary’s Street, Boston, Massachusetts 02215, USA\\
    $^2$Department of Physics, Boston University, 590 Commonwealth Avenue, Boston, Massachusetts 02215, USA\\
    $^3$U.S Army Combat Capabilities Development Command (DEVCOM) Army Research Laboratory (ARL), 2800 Powder Mill Road, Adelphi, MD 20783-7291 USA\\
    $^4$Division of Materials Science and Engineering, Boston University, 15 Saint Mary’s street, Brookline, Massachusetts 02446,USA
    }\\[0.4cm]
    
    {\small $^*$ Email: dalnegro@bu.edu}
\end{center}
\vspace*{1cm}
\twocolumngrid 


\onecolumngrid

\section{MSWNet network architectures and parameters}
Since each design optimization utilized a different maximum height, the spatial frequencies of the optimal solution and preferred network architecture varied over the designs. Table~\ref{tab:network_params} showcases the network architectures and scale values for each of the designs. Each architecture used the sine and SiLU activation functions for the first and hidden layers, respectively.
\begin{table}[htbp]
\caption{\label{tab:network_params} MSWN network parameters for each design}
\centering
\begin{tabular}{cccc}
Singlets & Number of hidden layers & Number of neurons & Scale values \\
\hline
$2\pi$ & 4 & 128 & 1, 32, 64, 128, 256, 512\\
$4\pi$ & 4 & 256 & 64, 128, 256\\
$6\pi$ & 4 & 256 & 64, 256\\
Doublets & & &\\
\hline
$2\pi$ & 4 & 256 & 32, 64, 128, 256\\
$4\pi$ & 4 & 256 & 128, 256\\
$6\pi$ & 4 & 256 & 64, 256\\
\end{tabular}
\end{table}

\section{Inverse-designed doublet using $h_{max}$ corresponding to a $6\pi$ phase shift}
\begin{figure}[htbp]
    \centering
    \includegraphics[width=0.7\textwidth]{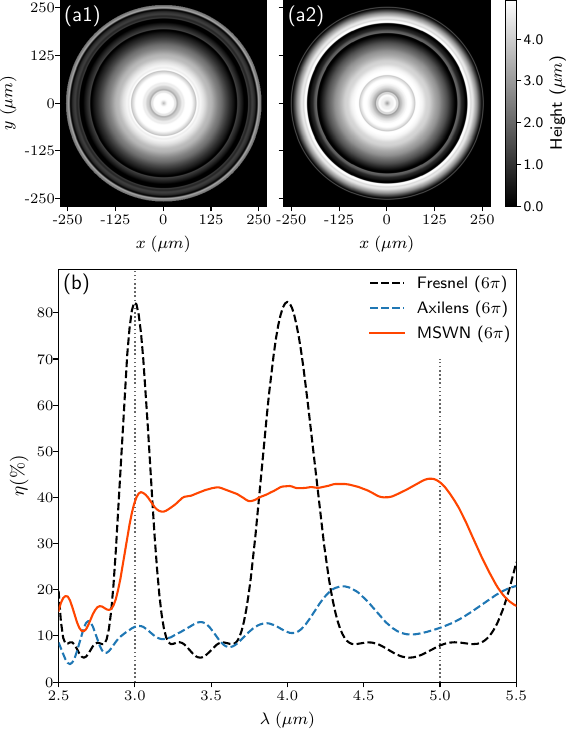}
    \caption{(a1,a2) MSWN-retrieved height profiles of the achromatic doublet utilizing a maximum height of 4.93 $\rm{\mu m}$ corresponding to a maximum phase shift of $6\pi$ at $\lambda=4 \rm{\;\mu m}$. (b) Simulated focusing efficiency $\eta$ as a function of wavelength of the MSWN-retrieved doublet (orange), axilens (dashed blue), and Fresnel lens (dashed black), using max heights determined by $6\pi$ phase shift at $\lambda=4 \rm{\;\mu m}$. The Fresnel lens is designed for $\lambda=12\rm{\;\mu m}$ displaying 2nd, 3rd, and 4th diffraction orders at $\lambda =6$, $4$, and $3\rm{\;\mu m}$, respectively. The target spectral range ($3-5\rm{\; \mu m}$) is indicated by the vertical gray lines.}
    \label{fig:supp_heights_efficiency_6pi}
\end{figure}

In the MSWNet optimization process, we found increasing the allowed height range bound by $0$ and $h_{max}$ greatly increased achromatic performance. Here, we analyze the performance of the optimized design using a maximum height to 4.93 $\rm{\;\mu m}$, corresponding to a total $6\pi$ phase shift at the central wavelength of $\lambda=4 \rm{\; \mu m}$ in silicon. Figure~\ref{fig:supp_heights_efficiency_6pi} displays the efficiency as a function of wavelength for the Fresnel lens (dashed black), axilens (dashed blue), and the inverse-designed doublet (orange). The Fresnel lens and axilens are designed using the same parameters in the main text, except with a maximum height of 4.93 $\rm{\;\mu m}$ for comparison.
Figure~\ref{fig:supp_heights_efficiency_6pi} clearly demonstrates that, even when compared with these thicker analytical designs, the inverse-designed doublet achieves a flatter efficiency profile and a superior average efficiency.

\begin{figure}[htbp]
    \centering
    \includegraphics[width=0.9\textwidth]{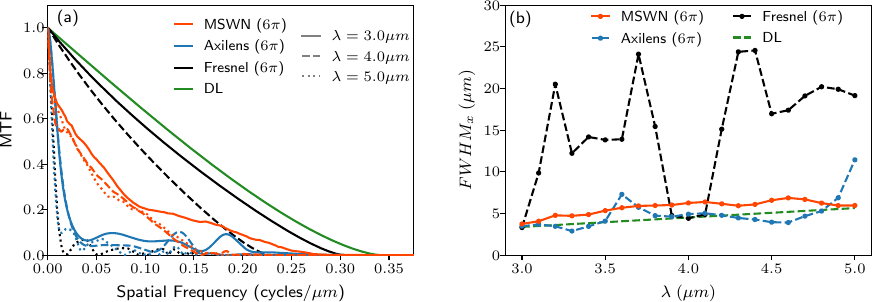}
    \caption{Performance metrics of lens designs. (a) Modulation-transfer function (MTF) of the inverse-designed lens (orange) compared with the axilens (blue), Fresnel lens (black) and diffraction-limited (green) with wavelength indicated by the solid, dashed, and dotted lines. (b) Full-width at half-maximums of the designs compared to the diffraction limited.}
    \label{fig:supp_MTF_FWHM_6pi}
\end{figure}

Figure~\ref{fig:supp_MTF_FWHM_6pi}(a) displays the modulation-transfer function (MTF) of the Fresnel lens (black), axilens (blue), and inverse-designed doublet (orange) at three wavelengths depicted by the solid, dashed and dotted lines corresponding to $\lambda = 3$, $4$, and $\rm{5\;\mu m}$ respectively. The diffraction-limited MTF at a wavelength of $3 \rm{\;\mu m}$ is shown in green for reference.
Here we can see the Fresnel lens exhibits near-diffraction-limited performance at $3$ and $4 \rm{\;\mu m}$ due to the diffracted orders at its design wavelength of $12 \rm{\;\mu m}$, with little resolution at $5 \rm{\;\mu m}$. However, this performance is expected to drop significantly at intermediate wavelengths, as seen from the efficiency curve in Fig.~\ref{fig:supp_heights_efficiency_6pi}(c). The axilens maintains little spatial frequencies at each of the analyzed wavelengths and is outperformed by the inverse-designed doublet, which exhibits the broadest frequency response.
Figure~\ref{fig:supp_MTF_FWHM_6pi}(b) displays the full-width at half-maximum (FWHM) values, as fit by an Airy pattern, for the three devices, compared with the diffraction limited in green. Once again, we can see small FWHM values for the Fresnel lens at its harmonic wavelengths of $3$ and $5 \rm{\;\mu m}$, exhibiting significant jumps in-between.
The axilens seems to have below diffraction-limited values across the spectrum, but this is due to the numerical fitting routine, which extracts the FWHM of the central lobe, neglecting the outer rings where intensity is distributed and lost. The inverse-designed doublet displays near-diffraction-limited values across the target wavelength spectrum, but with an improved PSF quality over the axilens. This is more clearly in Sec. S4, where we plot several of the fits for the designs.

\begin{figure}[htbp]
    \centering
    \includegraphics[width=0.9\textwidth]{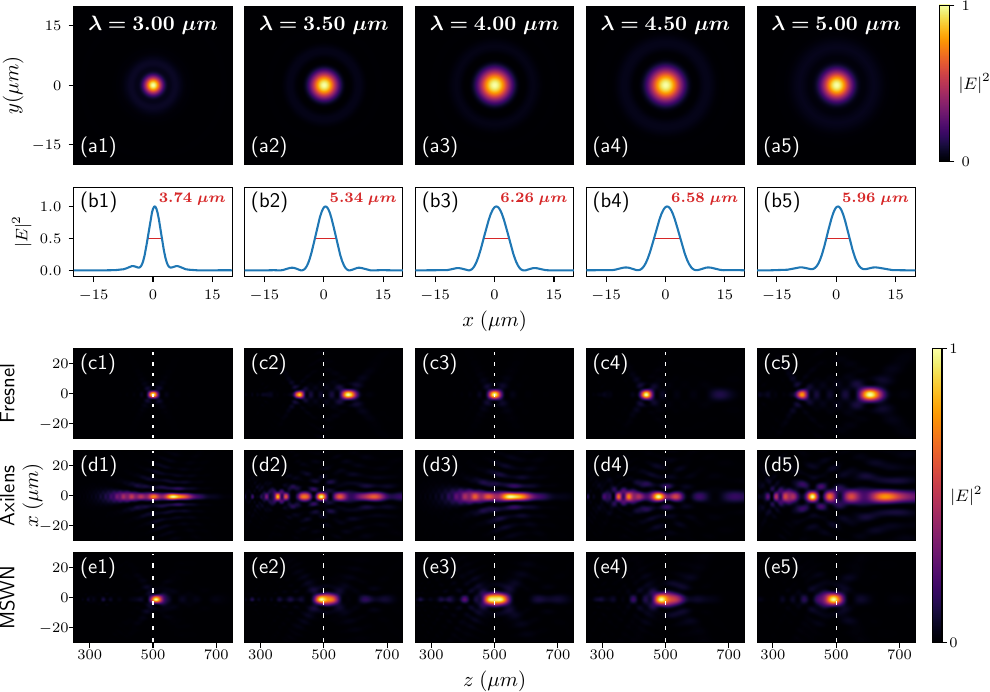}
    \caption{Wavelength-dependent focal characterization of the inverse-designed achromatic lens with a maximum height corresponding to a $6\pi$ phase-shift at $\lambda=4\rm{\;\mu m}$. (a1)--(a5) Normalized 2D transverse intensity profiles $|E|^2$ of the inverse-designed achromatic doublet at the designed focal plane ($f=500 \rm{\;\mu m}$) across the $3-5 \rm{\;\mu m}$ wavelength spectrum. (b1)--(b5) Corresponding 1D horizontal line-cuts ($y=0$). Full-width at half-maximums (FWHMs) fit are indicated by the red bounds with values shown in the top right of each panel. (c1)--(c5) to (e1)--(e5) Cross-sectional longitudinal intensity profiles along the propagation direction for: (c1)--(c5) a conventional Fresnel lens, (d1)--(d5) an axilens, and (e1)--(e5) the MSWN inverse-designed doublet. The vertical dotted white lines indicate the target focal length ($f=500 \rm{\;\mu m}$). Intensity values for all 2D and longitudinal plots are independently normalized to their respective global maxima.}
    \label{fig:supp_PSFs}
\end{figure}

Furthermore, Fig. \ref{fig:supp_PSFs} displays the normalized intensity profiles of: (a1)--(a5) the normalized transverse profiles of the inverse-designed $6\pi$ doublet, (b1)--(b5) the corresponding line-cuts with FWHMs indicated in red, (c1)--(c5) longitudinal intensity profiles of the Fresnel lens design, (d1)--(d5) longitudinal intensity profiles of the axilens design, and (e1)--(e5) longitudinal intensity profiles of the MSWNet inverse-designed doublet.
We can see the ultra-broadband performance of the $6\pi$ doublet with small traverse intensity profiles and a confined longitudinal profile across the $3-5 \rm{\;\mu m}$ spectrum.
Once again, the Fresnel lens exhibits significant focusing at $3$ and $4 \rm{\;\mu m}$ due to the 3rd and 4th diffracted orders of the design, with substantial chromatic shift in the focal position at $3.5 \rm{\; \mu m}$.

\section{Validity of scalar diffraction theory and analysis of feature sizes}

\begin{figure}[htbp]
    \centering
    \includegraphics[width=0.9\textwidth]{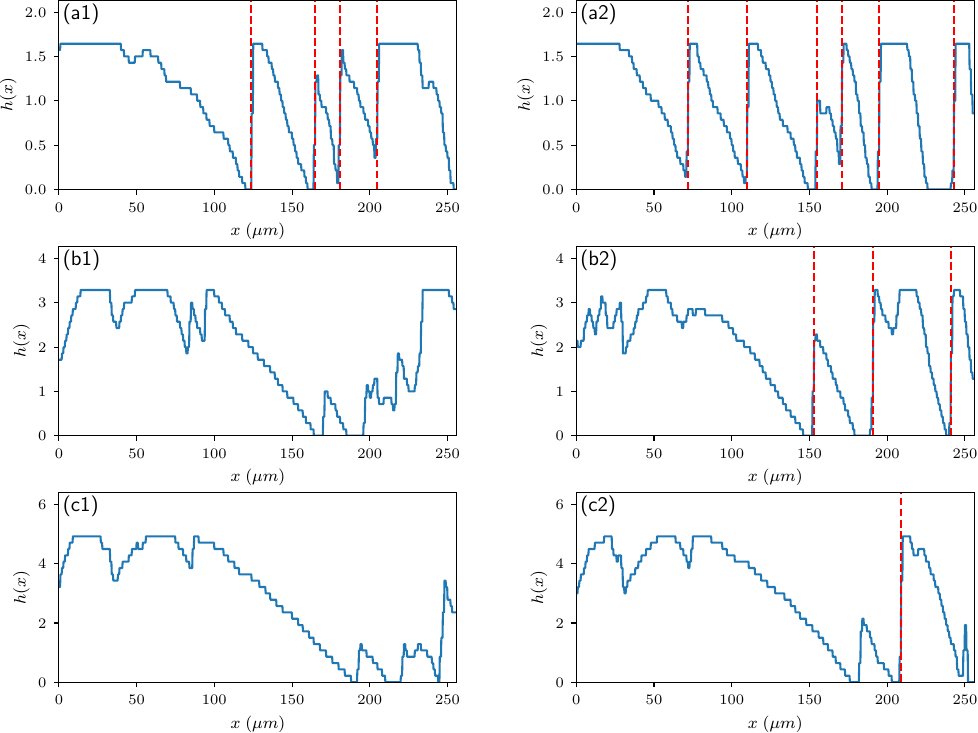}
    \caption{Structure analysis of inverse-designed diffractive doublets. Doublet layers using a max height of (a1, a2) 1.64 $\rm{\;\mu m}$, (b1, b2) 3.29 $\rm{\;\mu m}$, and (c1, c2) 4.93 $\rm{\;\mu m}$, discretized into 24 height levels. The horizontal red lines indicate areas where the height profiles exhibit abrupt changes.}
    \label{fig:supp_feature_sizes}
\end{figure}

Scalar diffraction theory is only accurate so long as the thin-element-approximation (TEA) holds, such that a diffractive layer is thin enough to act as a single phase transition, neglecting vector interactions of the electric field. It has been noted that this scalar approximation breaks down as the feature size approaches several wavelengths \cite{goodman2017introduction}.
However, in 2001, Mellin and Nordin demonstrated with (simulated) gratings etched into silicon, that the angular spectrum scalar theory remains accurate so long as the period remains larger than the free space wavelength \cite{mellin_limits_2001}. Here, they mention the pixel size is $\delta x=0.2\lambda$ ($\delta x=1 \rm{\;\mu m}$ and $\lambda=5\rm{\;\mu m}$), same as in our main text.
In their argument, the scalar theory depends less on the pixel size used, and more on the periodicity of the physical structure. Mellin and Nordin define ``zone sizes'', or the local period, as the region where the profile remains quasi-continuous between adjacent minimum features.
Here, we analyze the structure and periodic behavior of our inverse-designed doublet structures, to ensure they are properly modeled within the Rayleigh-Sommerfeld diffraction formulation. Figure \ref{fig:supp_feature_sizes} displays line-cuts of the height profiles for the designs using maximum phase shifts of $2\pi$, $4\pi$, and $6\pi$ In our designs, we found the smallest zone width among the doublet profiles to be 16 $\rm{\mu m}$ (Fig. \ref{fig:supp_feature_sizes}(a2)), falling within this limit. In each doublet layer, the minimum transverse feature size is $1\rm{\;\mu m}$, while the average feature size varies from $2$ to $3 \rm{\; \mu m}$. 

Another constraint for TEA to hold is the lateral displacement of the diffractive wave, $\delta d$ must be much less than the local zone size $\Delta x$ , i.e., $\delta d\ll \Delta x$, for the waves to not pass through multiple zones as it propagates. The lateral displacement can be written using the diffracted angle as $\delta d=h\tan{\theta}$. From the grating equation, considering only the first diffraction order, we have $n\sin{\theta}=\lambda/\Delta x$. Solving for the angle and substituting into the condition we obtain an expression for the height
\begin{equation}
    h_{max} \ll\frac{\Delta x\sqrt{(n\Delta x)^2 - \lambda^2}}{\lambda}
\end{equation}
In our case, using a wavelength of $\lambda=5 \mathrm{\;\mu m}$, minimum zone size $\Delta x=16 \mathrm{\;\mu m}$, and an index of $n=3.4$, we obtain a maximum height limit of $h_{max}=173 \mathrm{\;\mu m}$, falling well within this limit for our designs.

\section{Full-width at half-maximum estimation with Airy fitting}
\begin{figure}[htbp]
    \centering
    \includegraphics[width=0.9\textwidth]{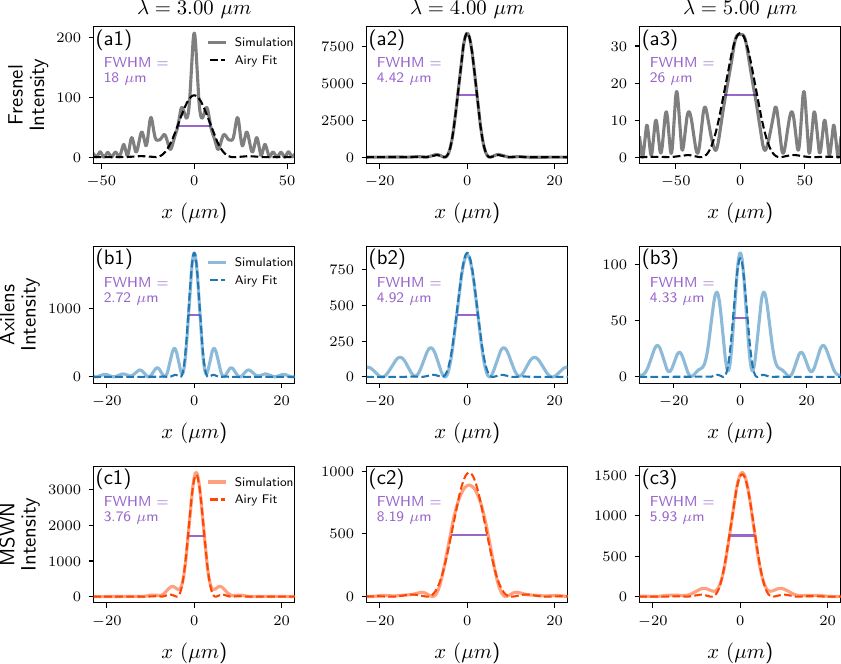}
    \caption{Horizontal line-cuts and corresponding Airy pattern fits of (a1--a3) Fresnel lens, (b1--b3) axilens, and (c1--c3) the inverse-designed doublet using $\phi_{max}=4\pi$, with columns corresponding to $\lambda = 3$, $4$, and $5\mathrm{\;\mu m}$. FWHMs are plotted as a horizontal line for visualization with the value indicated in purple.}
    \label{fig:supp_airy_fits}
\end{figure}

\begin{figure}[htbp]
    \centering
    \includegraphics[width=0.9\textwidth]{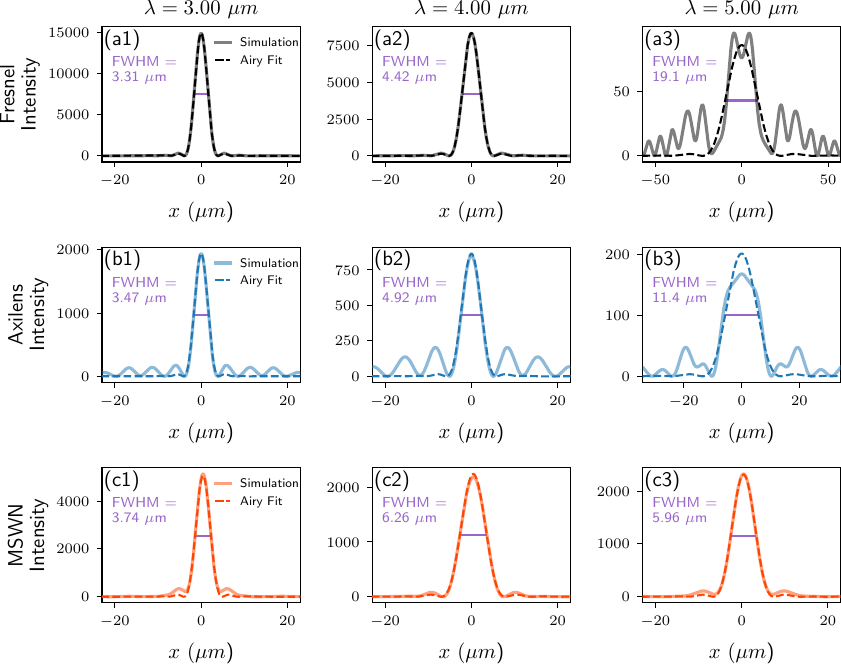}
    \caption{Horizontal line-cuts and corresponding Airy pattern fits of (a1--a3) Fresnel lens, (b1--b3) axilens, and (c1--c3) the inverse-designed doublet using $\phi_{max}=6\pi$, with columns corresponding to $\lambda = 3$, $4$, and $5\mathrm{\;\mu m}$. FWHMs are plotted as a horizontal line for visualization with the value indicated in purple.}
    \label{fig:supp_airy_fits_6pi}
\end{figure}

The FWHMs of each design was estimated through fitting the output intensities in the focal plane with an Airy function of the form:
\begin{equation}
    I(r)=A\left[2 \frac{J_1(wr)}{wr} \right]^2
\end{equation}

and the FWHM can be found from the zeros of the Bessel function as $\mathrm{FWHM}=3.232/w$ where $w$ is the fitted width. Figure~\ref{fig:supp_airy_fits} displays horizontal line-cuts of the output intensities and the corresponding fits for each of the $\phi_{max}=4\pi$ designs. Figures~\ref{fig:supp_airy_fits}(a1)--(a3) show the Fresnel lens output intensity and fits at 3,4 and 5$\mathrm{\;\mu m}$. Panels (b1)--(b3) display the results for the axilens while (c1)--(c3) show our MSWN inverse-designed doublet. From these line-cuts it is evident that the conventional Fresnel lens and axilens exhibit poor focal quality with higher intensity distributed in the side lobes. Additionally, as in Fig.~\ref{fig:supp_airy_fits}(b1) and (b3), the extracted FWHM of the axilens may remain artificially small due to the fitting routine capturing the main central lobe. The fits for the $\phi_{max}=6\pi$ devices presented in Sec. S2 are shown in Fig. \ref{fig:supp_airy_fits_6pi}. An improved PSF is seen at $\lambda=3 \mathrm{\;\mu m}$ for the Fresnel lens due to its 4th diffraction order of the designed wavelength at $\lambda=12 \mathrm{\;\mu m}$. Once again, the PSF of the inverse-designed achromatic doublet is superior and exhibits greater stability as a function of wavelength.

\end{document}